\documentclass[preprint,12pt]{elsarticle}

\usepackage{graphicx}
\usepackage{amssymb}

\usepackage{lineno}
\usepackage{color}

\begin{document}

\begin{frontmatter}


\title{A multi-grid Cellular Automaton model for simulating dendrite growth and its application in additive manufacturing}



\author[1,2]{Yefeng Yu}
\author[1]{Yang Li}
\author[1,3]{Feng Lin}
\author[2,3]{Wentao Yan}

\address[1]{Department of Mechanical Engineering, Tsinghua University, Beijing, China}
\address[2]{Department of Mechanical Engineering, National University of Singapore, Singapore}
\address[3]{Corresponding author: mpeyanw@nus.edu.sg, linfeng@tsinghua.edu.cn}

\begin{abstract}
The dendrite growth in casting and additive manufacturing is rather important and related to the formation of some defects. However, quantitatively simulating the growth of dendrites with arbitrary crystallographic orientations in 3-dimension(3D) is still very challenging. In the present work, we develop a multi-grid Cellular Automaton (CA) model for the dendrite growth. In this model, the interfacial area is further discretized into a child grid, on which the decentered octahedron growth algorithm is performed. The model is comprehensively and quantitatively verified by comparing with the prediction of analytical models and a published x-ray imaging observation result, proving that the model is quantitatively and morphologically accurate. After that, with the temperature gradient and cooling rate extracted from a finite-volume-method(FVM)-based thermal-fluid model, the model was applied in reproducing the dendrite growth process of nickel-based superalloy during a single-track electron beam melting process. The simulation results agree fairly well with the experimental observation, demonstrating the feasibility and effectiveness of using the model in additive manufacturing.

\end{abstract}


\end{frontmatter}


\section{Introduction}
\label{S:introduction}
Dendrite growth in casting and additive manufacturing has been proved related to some critical phenomena including the formation of pores \cite{plancher2019tracking}, initialization and propagation of cracks \cite{chauvet2018hot}, and precipitation of second phases \cite{tao2019crystal}, which play important roles in determining the mechanical properties of as-built products. In the past decades, dendrite growth, branching and fragmentation have been captured and investigated using x-ray imaging technology \cite{plancher2019tracking,yasuda2019dendrite,liotti2016spatial,cai20164d}. Nevertheless, only limited information was obtained as most of observations were two-dimensional (2D), and not enough to uncover the complex physical mechanisms in the actual 3-dimensional (3D) dendrite growth process. Numerical modeling, on the other hand,  has been demonstrated a powerful tool for understanding the dendrite growth process. 

To date, many efforts have been devoted in developing dendrite growth models using the Phase field (PF) and Cellular Automaton (CA) methods. The approach of representing the solid-liquid interface is one of their fundamental differences. In the phase field models, the solid-liquid interface is not explicitly tracked but described by a transition of a phase-field variable over the thickness of several grids. Therefore, to simulate the dendrite growth with a desirable resolution, a small mesh size is required, which will lead to a high computational cost. On the contrary, the solid-liquid interface is represented by a single layer of cells in the CA models. Therefore, coarse mesh can be used and thus much less computational power is demanded \cite{reuther2014perspectives}, which naturally enables the solution of the large-scale problems mentioned above. 

For the CA models, especially for 3D CA models, to quantitatively reproduce the growth of dendrites with arbitrary crystallographic orientations is never a trivial problem. In the CA models, the dendrite grows with the advancement of the solid-liquid interface, which is implicitly represented by the evolution of solid fraction in the interfacial cells. Furthermore, according to whether an interface is assumed inside an interfacial cell, the existing models can be divided into two categories as follows. 

For models in the first category, there is no necessity to determine the actual distribution of liquid and solid in a cell, i.e. the solid-liquid interface inside a cell, and the change of solid fraction can be accurately calculated by balancing the discrepancy between the local concentration and the equilibrium concentration. Therefore, these models are usually quantitative \cite{reuther2014perspectives} and match well with the prediction of the Lipton-Glicksman-Kurz (L-G-K) model \cite{pan2010three}. However, the serious anisotropy from the nature of the grid makes it hard to achieve crystallographic orientations other than along the x, y and z axes. Nevertheless, by introducing an orientation-related curvature when calculating the driving force of dendrite growth, some researchers succeeded in achieving arbitrary orientations. For instance, \citet{pan2010three} adopted a weighted mean curvature when deriving the equation of the equilibrium concentration, which is the critical part in their formulations of the growth kinetics. The method and similar approaches are also utilized by other researcher \cite{chen2015cellular,eshraghi2013three,choudhury2012comparison,jelinek2014large,zaeem2013modeling}. However, the key point of these methods to achieve arbitrary orientations is to ensure the curvature undercooling be more predominant than the thermal undercooling (see Eq.\ \ref{eq:cleq}), which requires an elaborate curvature evaluation. Therefore, a sophisticated curvature calculation algorithm and dense mesh are necessary, thereby increasing the computational cost greatly.

As for models belonging to the second category, a virtual interface in the interfacial cell is assumed, and its velocity is determined by the deviation of the interfacial concentration from the equilibrium value. The change of the solid fraction is then obtained by evaluating the area (or volume) swept by the interface. The adoptions of the decentered square growth algorithm in 2D models and the decentered octahedron growth algorithm \cite{rappaz1993probabilistic} in 3D models make the simulation of the growth of dendrites with arbitrary orientations feasible \cite{nakagawa2006dendrite,chen2014modified,wang2003model} even without sophisticated curvature calculation algorithm and dense mesh. However, due to the complex interface consisting of squares and octahedrons, it is rather difficult to accurately evaluate the change of the solid fraction and thus challenging to achieve accurate quantitative simulation, especially in 3D. Therefore, only a few 3D dendrite growth models were developed based on this method \cite{wang2003model,2020Multi,2016A}, and none of them is demonstrated quantitatively accurate and able to simulate the growth of dendrite with arbitrary orientations. 

In this work, a 3D-CA model enhanced by a multi-grid method is developed to simulate the dendrite growth. The decentered octahedron growth algorithm is adopted and performed on the child grid that only exists at the solid-liquid interface area. With this method, the arbitrary orientations of dendrites can be easily achieved without a sophisticated curvature calculation algorithm, and the change of the solid fraction in interfacial cells can be accurately evaluated. The model is comprehensively and quantitatively verified by comparing with the prediction of analytical models, and then validated against a published x-ray imaging observation \cite{yasuda2019dendrite}. 

Dendrite growth is proven critical in the formation of precipitates and cracks during the AM process \cite{chauvet2018hot,zhou2018causes,lu2020hot}. With the phase field models and CA models, researchers have successfully reproduced the 2D dendrite growth process under the AM conditions, and studied the impact of solidification condition on the microstructure evolution \cite{nie2014numerical,wu2018phase,gong2015phase,wang2019investigation}. However, to date, only very few works try to simulate the dendrite growth in 3D \cite{sun2019gpu}. No work attempts to adopted 3D CA-based models, which consume much less computational cost than the phase field models, and are believed more suitable for large scale problems. Therefore, in this study, the 3D dendrite growth of Ni-Nb alloy in the single-track electron beam melting process is simulated using the present model. The simulation result is then compared with experimental results, thereby demonstrating the model's feasibility and effectiveness in the investigation of microstructure evolution during the additive manufacturing process. 

\section{Model description}
\label{S:model}
\subsection{Multi-grid method}
\label{S:m_g}
\begin{figure}[htp]
    \centering
    \includegraphics[width=0.8\linewidth]{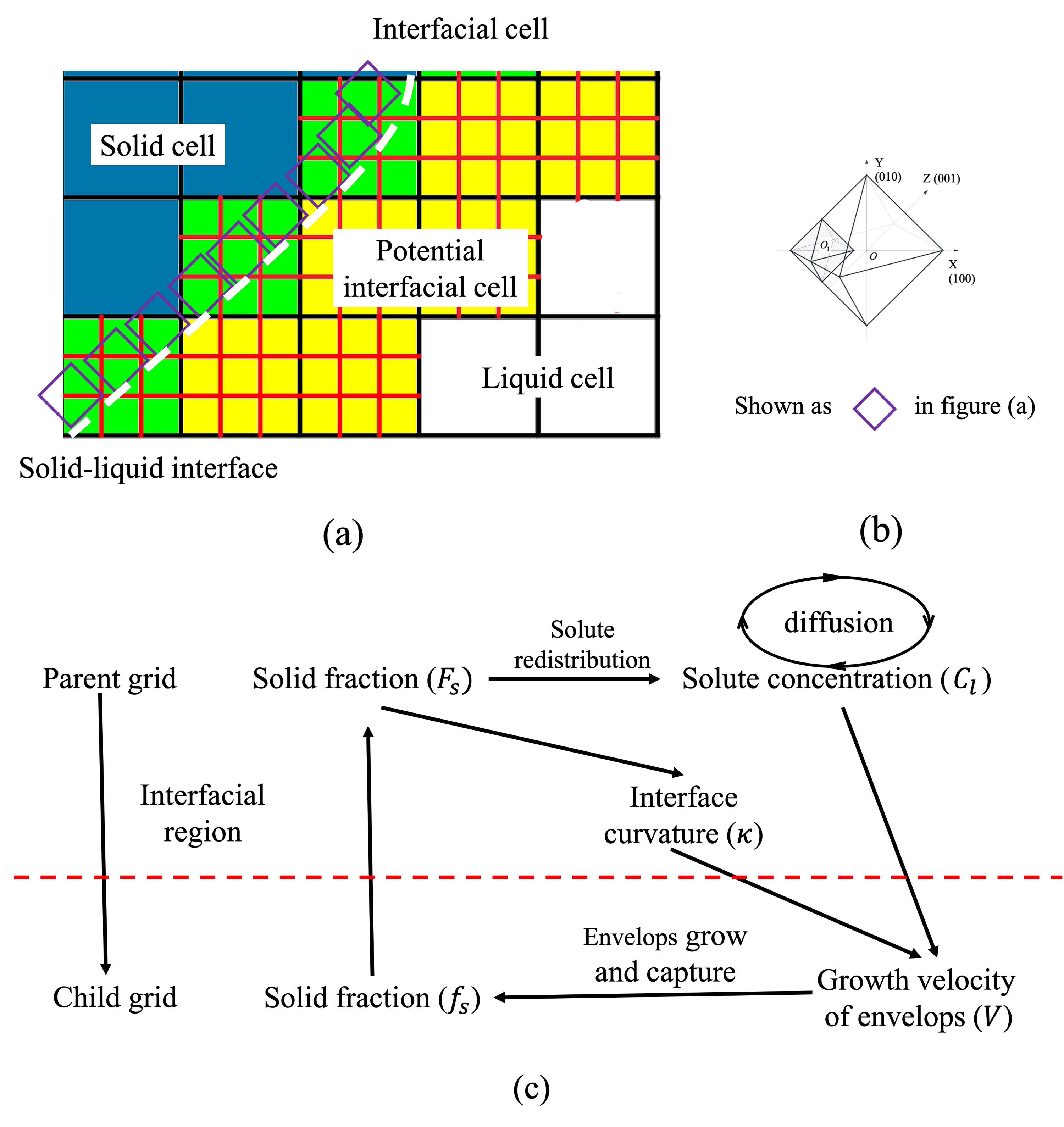}
    \caption{(a) Two dimensional schematic of the multi-grid structure; (b) Schematic diagram of decentered octahedron growth algorithm. The octahedral envelopes are shown as violet squares in Fig. (a). These envelopes consist of the solid-liquid interface; (c) Schematic diagram of the "handshaking" between the parent grid and the child grid.}
    \label{fig:algorithm}
\end{figure}

 Fig.\ \ref{fig:algorithm}a illustrates the multi-grid method in 2D for simplicity. First of all, similar to the conventional CA models, the calculation domain is discretized into uniform cubic cells, i.e. the coarse grid (hereafter called as parent grid), and the cell size is $d_{CELL}$. In the parent grid, each cell is associated with a set of variables ($F_s$, $C_l$, $g$, $S$): $F_s$ is the solid fraction of this cell, $C_l$ is the solute concentration in the liquid, $g$ is the identification number of the dendrite, to which the cell belongs, and $S$ is the status of this cell ($S=-1$: solidified cell; $S=0$: interfacial cell; $S=1$: liquid cell; $S=2$: potential interfacial cell). The interfacial cells (green cells in Fig.\ \ref{fig:algorithm}a) and potential interfacial cells (yellow cells in Fig.\ \ref{fig:algorithm}a) are tracked and further discretized into smaller uniform cubic cells (hereafter called as child grid). The cell size of the child grid $d_{cell}$ equals to $d_{CELL}/n_{sub}$, and $n_{sub}$ is an odd number bigger than 1 in this study to guarantee the symmetry of the simulated dendrite structure. That is, there are $n_{sub}^3$ child-grid cells in a parent grid cell. Similarly, there are several variables associated with the child-grid cell, they are $f_s$, the solid fraction, and $s$, the status of the cell ($s=0$: solid, and $s=1$: liquid). 
 
In this model, the decentered octahedron growth algorithm (Fig.\ \ref{fig:algorithm}b) is implemented on the child grid to reproduce the movement of the solid-liquid interface. As Fig.\ \ref{fig:algorithm}a shows, the octahedral envelopes are located at the interfacial cells and constitute the solid-liquid interface (see the white dashed line). The six half-diagonals of these octahedral envelopes correspond to the crystallographic orientation of the dendrite structure, which implicitly incorporates the anisotropic surface energy. More details of this algorithm can be referred to Rappaz \cite{rappaz1993probabilistic} and will not be repeated here. During the simulation, the child grid will be updated with the advancement of the solid-liquid interface: (1) the child grid in the parent-grid cell, which has fully solidified and has no envelopes, is removed; (2) the potential interfacial cell turns into an interfacial cell once at least one of its child-grid cells is captured; (3) new child grid will be generated at the liquid cells that are adjacent to the new interfacial cell.
 
The ``handshaking'' between the parent grid and the child grid is illustrated in Fig.\ \ref{fig:algorithm}c, where algorithms above the red dashed line are exerted on the parent grid, and those below the line are applied on the child grid. During the simulation, (1) the octahedral envelopes grow up by the movement calculated according to the interfacial solute concentration ($C_l$) and interface curvature ($\kappa$) at the parent grid scale; (2) envelopes capture their neighboring cells, updating the solid fraction ($f_s$) in the child grid; (3) the solid fraction ($F_s$) and the solute concentration ($C_l$) in the parent grid are updated; (4) the diffusion is calculated at the parent grid. These four procedures link the parent grid with the child grid and would be repeated till the end of the simulation. The detailed algorithms are introduced in the following sections.

\subsection{Growth kinetics}
In this model, the driving force for the solid-liquid interface movement is the deviation of the local liquid composition ($C_l$) from the equilibrium composition ($C_l^{eq}$). Within one time step ($\Delta t$), the change of solid fraction $\Delta F_s$ in the parent-grid cell can be calculated as:
\begin{equation}
    \Delta F_s=\frac{(1-F_s)(C_l^{eq}-C_l)}{C_l^{eq}(1-k)}
    \label{eq:dFs}
\end{equation}
\begin{equation}
    C_l^{eq}=C_0+\frac{T-T_0^{eq}}{m_l}+\frac{\Gamma \cdot \kappa}{m_l}
    \label{eq:cleq}
\end{equation}
where $k$ is the solute partition coefficient, $C_0$ is the initial solute concentration, $T_0^{eq}$ is the equilibrium temperature for the initial composition, $T$ is the local temperature, $\Gamma$ is the Gibbs-Thomson coefficient, $m_l$ is the liquidus slope, and $\kappa$ is the interface curvature and calculated using the counting-cell method \cite{kremeyer1998cellular} in the parent grid. The 3D version of this method is used in the present model as Eq.\ \ref{eq:kappa}.
\begin{equation}
    \kappa=\frac{\alpha}{d_{CELL}}\frac{51-\sum{F_s}}{81}
    \label{eq:kappa}
\end{equation}
where 81 is the number of cells with the distance from the interfacial cell being within $\sqrt{6} d_{CELL}$, $\sum F_s$ is the summed solid fraction of cells that belong to the same dendrite as the interfacial cell do among the 81 neighboring cells, 51 is the summed solid fraction when the interface is flat, i.e. $\kappa=0$, and $\alpha$ is a shape factor equal to 4.3899, which is derived by regressing the calculated radii ($1/{\kappa}$) on the predefined sphere radii.

Thus, the interface movement ($\Delta s$) is:
\begin{equation}
    \Delta s=\Delta F_s \cdot d_{CELL}
    \label{eq:ds}
\end{equation}

In this model, the tips' movement of all envelopes in the parent-grid cell is defined to be the same as the interface movement calculated on the parent grid, i.e., $\Delta s$. In the end of this time step, the envelope size $L$ which is the distance from the envelope center to its surface becomes:
\begin{equation}
    L=L_0+\frac{\Delta s}{\sqrt{3}}
    \label{eq:e_Length}
\end{equation}
where $L_0$ is the envelope size in the beginning of this time step.

\subsection{Calculation of solid fraction}

\begin{figure}[htp]
    \centering
    \includegraphics[width=\linewidth]{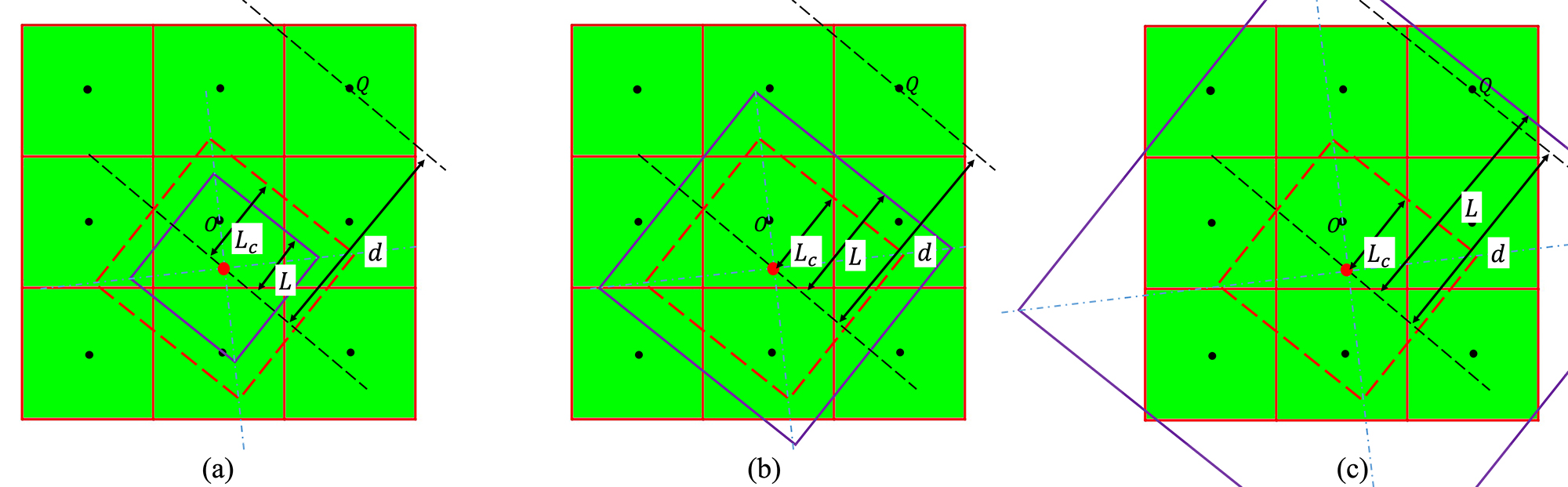}
    \caption{2D schematic diagram of the solid fraction ($f_s$) evaluation algorithm.}
    \label{fig:eva_solid}
\end{figure}

To simulate the evolution of the solid fraction in child-grid cells, in the present study, a solid fraction calculation method is designed and applied on the child grid. $d$ is the distance from the envelope center to the center of cell $Q$ along the normal direction of the envelope surface closest to cell $Q$, $L_c$ is the maximum size of new envelopes and equals to $\sqrt{2}d_{cell}$ according to the decentered octahedron growth algorithm. The solid fraction in cell $Q$ ($f_s^Q$) updates as follows:
\begin{itemize}
    \item[1)] $L<L_c$: as shown in Fig.\ \ref{fig:eva_solid}a, the envelope is too small to engulf cell $Q$, so $f_s^Q=0$.
    \item[2)] $L_c<L<d$: as shown in Fig.\ \ref{fig:eva_solid}b, cell $Q$ is outside of the envelope but partially engulfed by the envelope, the solid fraction will be:
    \begin{equation}
        f_s^Q=\frac{L-L_c}{d-L_c}
        \label{eq:fs}
    \end{equation}
    \item[3)] $L \ge d$: as shown in Fig.\ \ref{fig:eva_solid}c, cell $Q$ is captured by the envelope, then $f_s^Q=1$. At the same time, a new envelope hosted by cell $Q$ is generated according to the decentered octahedron growth algorithm.
\end{itemize}

If a cell is the neighboring cell of several envelopes, its solid fraction will be the maximum $f_s$ calculated as Eq.\ \ref{eq:fs}. After the completing the growth and capture process of all envelopes, the solid fraction in the parent-grid cells ($F_s$) can be obtained by averaging the solid fraction of its child-grid cells as Eq.\ \ref{eq:Fs}. 

\begin{equation}
    F_s=\frac{\sum f_s}{n_{sub}^3}
    \label{eq:Fs}
\end{equation}

As the dendrite grows up, the solute partition occurs at the interfacial cells. The solute concentration of the newly solidified child-grid cells will be $k \cdot C_l$. The rejected ($(1-k)C_l$ when $k<1$) or absorbed ($(k-1)C_l$ when $k>1$) is evenly added to or subtracted from the liquid part of this parent-grid cell and its neighboring parent-grid cells.

\subsection{Solute diffusion}
\label{S:solute_diff}
As shown in Fig.\ \ref{fig:algorithm}c, the solute diffusion model is defined on the parent grid. Usually, the diffusion coefficient in liquid is several orders of magnitude higher than that in solid, thus only diffusion in liquid is accounted here. In this study, the solute diffusion is modeled using the Fick's law:
\begin{equation}
    \frac{\partial C_l}{\partial t}=\nabla \cdot (D \nabla C_l)
    \label{eq:ficklaw}
\end{equation}
where $D$ is the effective diffusion coefficient and dependent on the cell status. In this model, the effective diffusion coefficient ($D_{AB}$) between cell $A$ and one of its Von-Neumann neighboring cell $B$ is defined as:
\begin{equation}
    D_{AB}=\frac{D_l^A +D_l^B}{2} \sqrt{(1-F_s^A) \cdot (1-F_s^B)}
    \label{eq:eff_D}
\end{equation}
where $D_l^A$ and $D_l^B$ are the diffusion coefficients in cell $A$ and cell $B$, respectively, $F_s^A$ and $F_s^B$ are the solid fractions in cell $A$ and cell $B$, respectively. Based on this equation, the diffusion coefficient is 0 in the fully solidified region, $D_l$ in the liquid area, and ranges from 0 to $D_l$ in the interfacial region.

\subsection{Model implementation}
The time step size ($\Delta t$) is predefined before the simulation starts and equals to $(d_{CELL})^2/(4D_l)$ in this study in order to guarantee the stability of the solute diffusion model. Within one time step, the calculation scheme is as follows:
\begin{itemize}
    \item [1)] Loop over all interfacial parent-grid cells and calculate the interface movements.
    \item[2)] Loop over all envelopes to complete the growth and capture processes and calculate the solid fraction in child-grid cells, then update the solid fraction in parent-grid cells. 
    \item[3)] Update the child grid as presented in Section \ref{S:m_g}.
    \item[4)] Loop over all parent-grid cells with solid fraction changed in this time step to update its solute concentration. 
    \item[5)] Calculate the diffusion and then update the temperature field if necessary.
\end{itemize}

\section{Results and discussion}
\label{S:R_d}
\subsection{Model verification}
The dendrite growth in Fe-0.6 wt.\%C alloy melt is simulated in this section. The critical properties of the material are listed on Table. \ref{tab:Fe_C_alloy}.

\begin{table}[htp]
    \centering
        \caption{Critical physical properties of Fe-C alloy \cite{pan2010three}}
    \begin{tabular}{c c}
    \hline
    Properties & Values\\
    \hline
       Solute diffusion coefficient in the liquid ($D_l$) &  $2\times 10^{-9}$  $\rm m^2/s$\\
       Solute diffusion coefficient in the solid ($D_s$) & $5\times 10^{-10}$ $\rm m^2/s$\\
        Liquidus slope ($m_l$) & -80 $\rm K/(wt. \%$)\\
        Partition coefficient ($k$) & 0.34\\
        Gibbs-Thomson coefficient ($\Gamma$) & $1.9\times 10^{-7}$ $\rm m\cdot K$ \\
        \hline
    \end{tabular}
    \label{tab:Fe_C_alloy}
\end{table}

\subsubsection{Multi-dendrite growth}

The present model's capability to reproduce the growth of dendrites with arbitrary crystallographic orientations is verified with a simulation of multi-dendrite growth in the Fe-0.6 wt.\%C alloy melt. The simulation domain is $150\times 150\times 150$ $\rm \mu m^3$, $d_{CELL}=1$ $\rm \mu m$ and $n_{sub}=3$. The undercooling $\Delta T$ is 8 K, and before the simulation, 27 dendrite seeds with randomly assigned orientations are randomly placed in the simulation domain.

\begin{figure}[htp]
    \centering
    \includegraphics[width=0.8\linewidth]{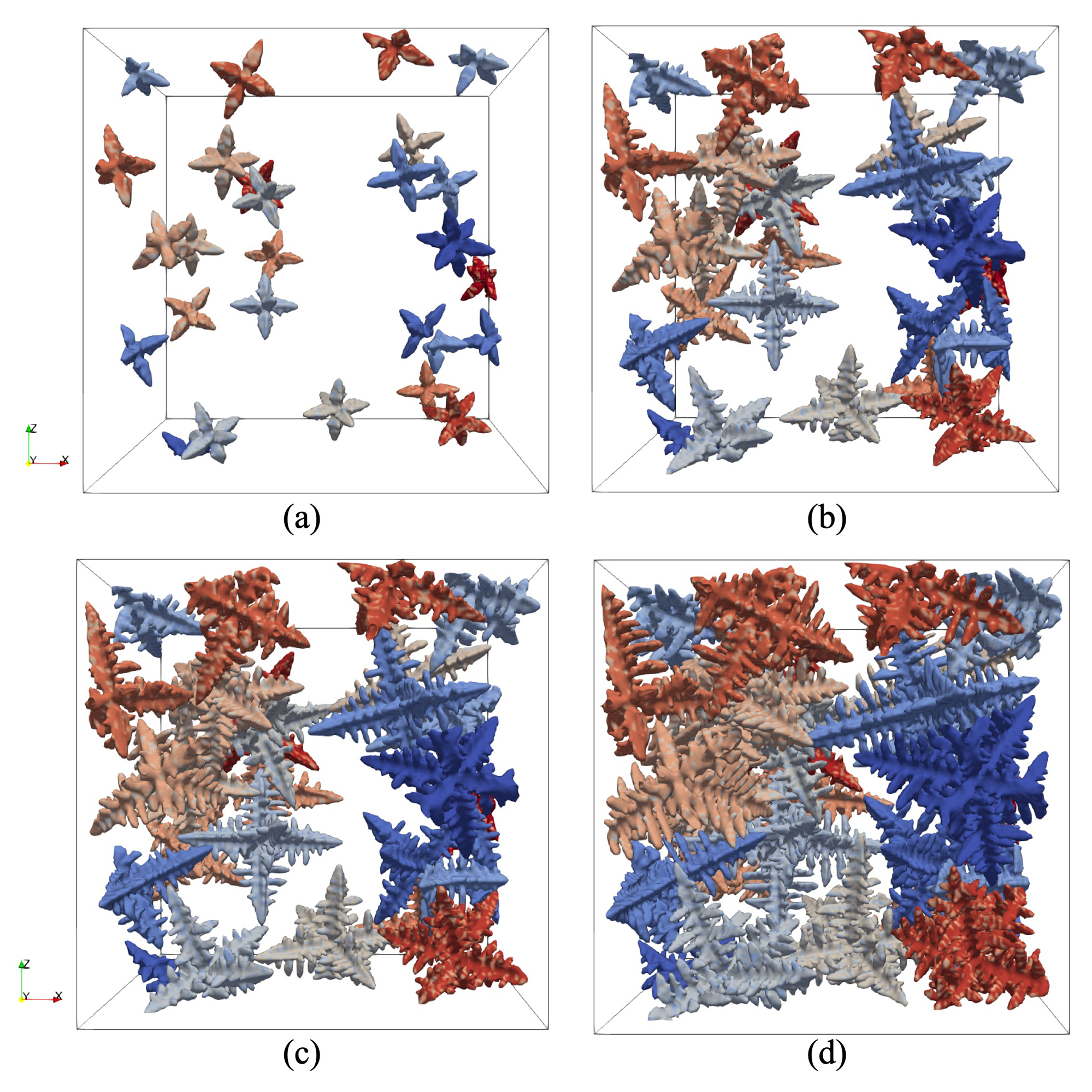}
    \caption{Simulation results of multi-dendrite growth at (a) $t=0.008$ s, (b) $t=0.016$ s, (c) $t=0.020$ s, (d) $t=0.028$ s.}
    \label{fig:multi_dendrite}
\end{figure}

Fig.\ \ref{fig:multi_dendrite} presents the simulated dendrite morphology in chronological order. Dendrites exhibit a branchless needle shape when their sizes are small and then develop the secondary and tertiary arms. As dendrites grow up, they start to influence each other, some dendritic trunks stop growing longer, and coarsening of trunks and arms becomes prevalent. Evidently, the growth of dendrites with arbitrary orientation is successfully reproduced with the present model.

\subsubsection{Comparison with the K-G-T model}
\label{S:L-G-K}
\begin{figure}[htp]
    \centering
    \includegraphics[width=0.6\linewidth]{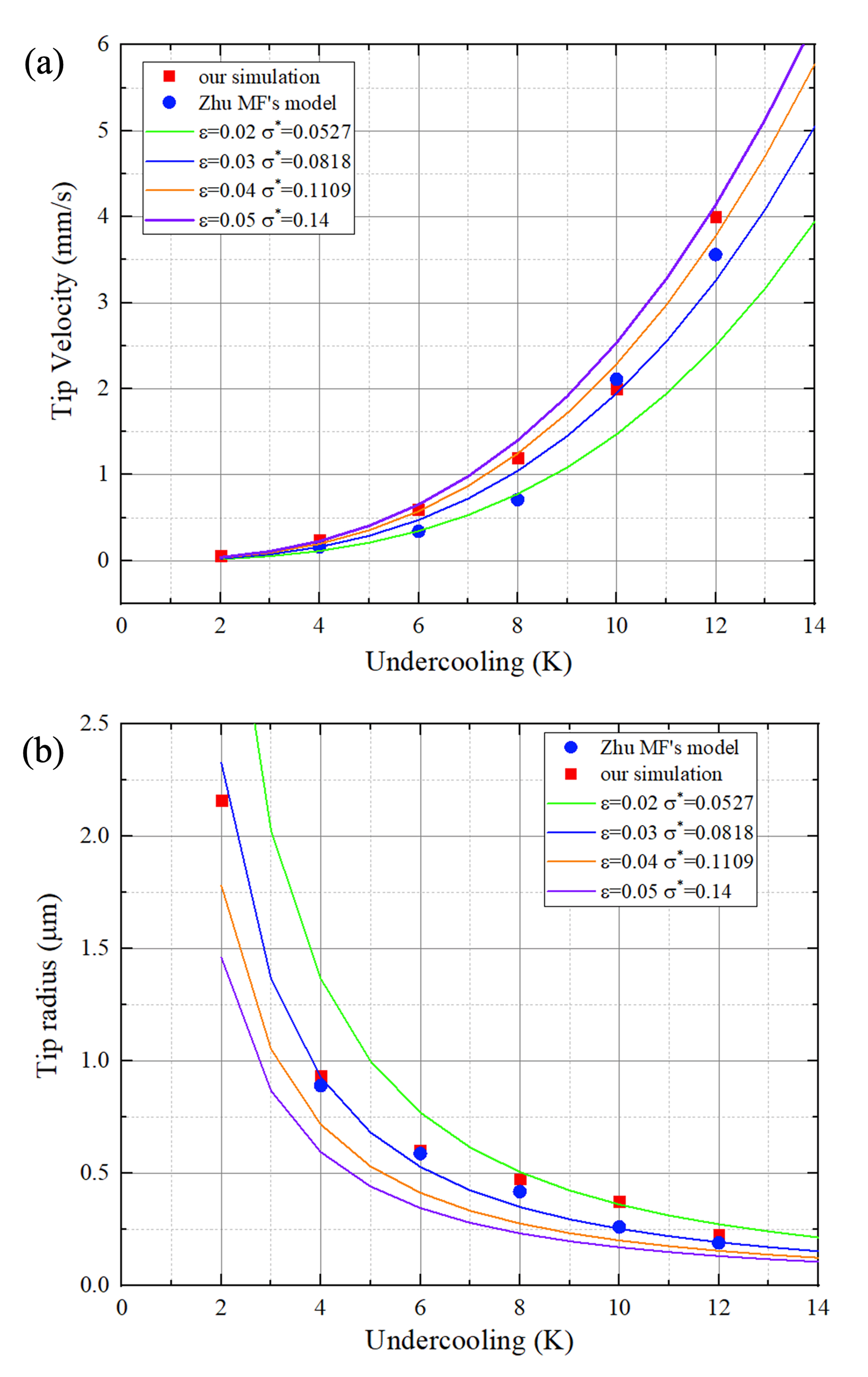}
    \caption{(a) Tip velocities calculated by our model (red squares), the K-G-T model with different $\sigma^*$ (solid curves) and Zhu MF's model \cite{pan2010three} (blue circles); (b) Tip radii calculated by our model (red squares), the K-G-T predictions with different $\sigma^*$ values (solid curves) and Zhu MF's model \cite{pan2010three} (blue circles).}
    \label{fig:L-G-K_comparison}
\end{figure}

In the L-G-K model \cite{lipton1987equiaxed}, for dendrite growth in the low Péclet number regime, it is assumed that the dendrite grows steadily into a melt of constant composition, and the mass and heat transport are merely influenced by diffusion during the process. In the present work, a simplified L-G-K model, i.e., the Kurz–Giovanola–
Trivedi(K-G-T) model\cite{kurz1986theory}, where the heat transport is not accounted, is used, i.e., K-G-T model. Therefore, the melt undercooling $\Delta T$, which drives the dendrite growth, is given as:
\begin{equation}
    \Delta T=\Delta T_c +\Delta T_R
    \label{eq:delta_T}
\end{equation}
where the $\Delta T_c$ is the solutal undercooling and $\Delta T_R$ is the curvature undercooling, given as:
\begin{equation}
    \Delta T_c=m_l\cdot C_0\cdot (1-\frac{1}{1-(1-k)\cdot I_v(P_c)})
    \label{eq:delta_TC}
\end{equation}
\begin{equation}
    \Delta T_R=\frac{2\Gamma}{R}
    \label{eq:delta_TR}
\end{equation}
where $P_c$ is the solutal Péclet number ($P_c=\frac{VR}{2D_l}$, $V$ is the tip velocity and $R$ is the tip radius. In this section, the calculated Péclet number ranges from 0.05 to 0.23), and $I_v(P_c)$ is the well-know Ivantsov function, which solves the transport problem near the dendrite tips, given as:
\begin{equation}
    I_v(P_c)=P_c\cdot exp(P_c) \int_{P_c}^{\infty} \frac{e^{-x}}{x} dx
    \label{eq:ivpc}
\end{equation}
In addition, under the criterion of marginal stability \cite{lipton1987equiaxed,kurz1989fundamental}, the curvature undercooling $\Delta T_R$ can be written as:
\begin{equation}
    \Delta T_R=-\frac{4\sigma ^* P_c m_l C_0 (1-k)}{1-(1-k)I_v(P_c)}
    \label{eq:dtR_2}
\end{equation}
where $\sigma^*$ is stability constant and affected by the degree of surface energy anisotropy ($\varepsilon$) according to the microsolvability theory \cite{barbieri1989predictions}. With Eq.\ \ref{eq:delta_T}, \ref{eq:delta_TC}, \ref{eq:delta_TR} and the criterion of marginal stability, the tip velocities $V$ and tip radii ($R$) at different undercoolings $\Delta T$ can be calculated. It is critical to note $\varepsilon$ is not explicitly incorporated in the present model. Therefore, in this study, $\sigma^*$ values corresponding to different $\varepsilon$ (0.02, 0.03, 0.04 and 0.05) are used to generate four set of predictions, which are presented in Fig.\ \ref{fig:L-G-K_comparison} with solid curves. 

Simulations of single dendrite growth at different undercoolings are conducted. The simulation domain is a $200\times 200\times 200$ mesh. The parent-grid cell size $d_{CELL}$ varies from 0.04 to 0.1 $\rm \mu m$ for simulations with different undercoolings in order to evaluate the curvature accurately. Additionally, $n_{sub}$ is 3, i.e., $d_{cell}=d_{CELL}/3$. At the beginning of the simulation, within the central parent-grid cell, a dendrite seed with the orientation along the x, y and z axes is placed at its central child-grid cell. 

\begin{figure}[htp]
    \centering
    \includegraphics[width=0.55\linewidth]{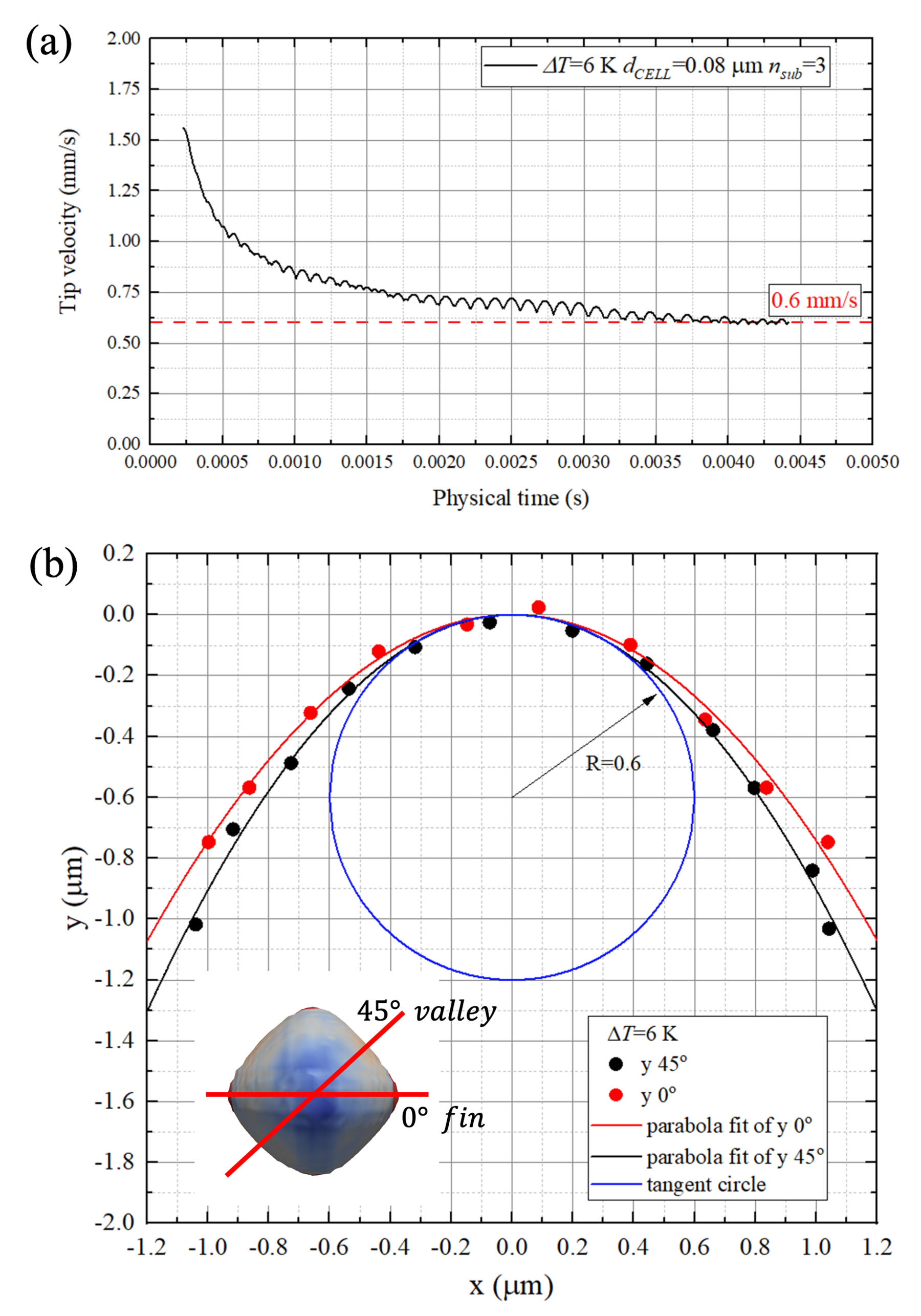}
    \caption{(a) Evolution of tip velocity in the simulation with $\Delta T=6$ K; (b) Methodology of tip radius measurement, the inserted figure shows the positions of $0^{\circ}$ plane and $45^{\circ}$ plane.}
    \label{fig:steady_state_r}
\end{figure}

The tip velocities and tip radii are measured when the dendrite reaches the status of steady growth. Fig.\ \ref{fig:steady_state_r}a shows the evolution of tip velocity in the simulation with $\Delta T=6$ K. The tip velocity is very high at the beginning, then decreases gradually, and finally becomes steady at 0.6 mm/s. Fig.\ \ref{fig:steady_state_r}b shows the methodology of tip radius measurement. The tip contours in the $0^{\circ}$ and $45^{\circ}$ planes (corresponding to the fin and valley of the dendrite tip as shown in the inserted figure) are extracted and fitted with parabolas. At $x=0$, the tangent circles for the two parabolas are calculated, and the tip radius ($R$) is calculated by averaging the radii of the tangent circles. The blue circle in Fig.\ \ref{fig:steady_state_r}b is the tangent circle with the calculated radius.

The simulated tip velocities and radii at different undercoolings are plotted in Fig.\ \ref{fig:L-G-K_comparison}a and b, accompanied by the calculated tip velocities and radii using the K-G-T model with different $\sigma^*$. Obviously, the simulation results show a fairly good agreement with the prediction of the K-G-T model. Furthermore, the simulated tip velocities lie between the blue curve ($\varepsilon=0.03$, $\sigma^*=0.0818$) and the violet curve ($\varepsilon=0.05$, $\sigma^*=0.14$), while the simulated tip radii lie between the orange curve ($\varepsilon=0.04$, $\sigma^*=0.01109$) and the green curve ($\varepsilon=0.02$, $\sigma^*=0.0517$). Therefore, when considering the distance between data points and the K-G-T curves in these two figures at the same time, it is reasonable to believe that $\varepsilon \approx 0.03$ for the presented model. \citet{pan2010three} also verified his model (first category in the introduction section) by comparing with the prediction of the K-G-T model. In his model, the degree of surface energy anisotropy is explicitly included in equations and equals to 0.03. Their results are presented in Fig.\ \ref{fig:L-G-K_comparison}a and b using blue solid circles. Apparently, our simulation results is quite similar with theirs.

\subsubsection{Tip morphology}
\begin{figure}[htp]
    \centering
    \includegraphics[width=0.5\linewidth]{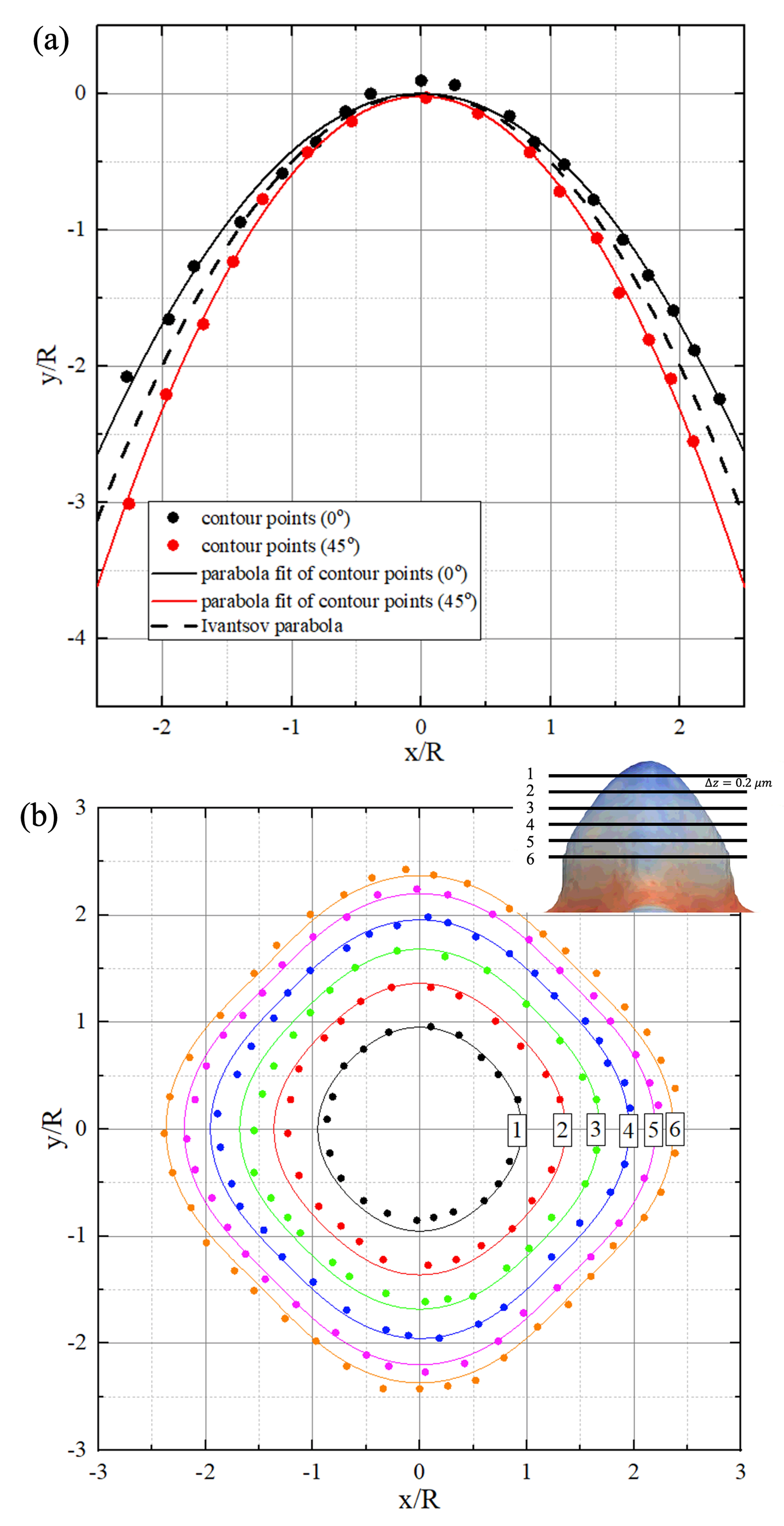}
    \caption{Steady-state tip morphology of the dendrite formed in the simulation with $\Delta T=8$ K: (a) longitudinal tip contours taken along fin ($0^{\circ}$, black solid circles) and valley ($45^{\circ}$,red solid circles), accompanied by the parabolic fitting curves and Ivantsov parabola ($y/R=-0.5(x/R)^2$). (b) Transverse sections of a simulated tip at different along the Z direction with a interval of 0.2 $\rm \mu m$ as shown in the inserted figure. The distance between the first section and the tip is also 0.2 $\rm \mu m$. In the figure, solid circles are contour points and solid curves are the shapes of the Fourier decomposition.}
    \label{fig:tip_morphology}
\end{figure}

In this section, the steady-state morphology of dendrite tip obtained when $\Delta T=8$ K is studied in detail. Fig.\ \ref{fig:tip_morphology}a shows the longitudinal cross sections of the dendrite tip. The horizontal and vertical tick labels are $x/R$ and $y/R$, where $R$ is the tip radius. The black and red solid circles are points on tip contours ($F_s=0.5$) along the fin ($0^{\circ}$ plane) and valley ($45^{\circ}$ plane), respectively. The contour points are fitted with two parabolas, which almost overlap near $x/R=0$ and separate when far away from $x/R=0$. This indicates that the shape of cross sections close to the tip is nearly axisymmetric, and become nonaxisymmetric gradually with the increase of the distance from the tip, which is consistent with the simulation results of the phase field model \cite{karma1998quantitative}, where an explicit surface energy anisotropy is adopted and incorporated in the free energy function.

Ivantsov developed a stationary solution to the free-growth model without surface tension \cite{ivantsov1952growth}, which describes a diffusion-controlled growth of a needle-shape crystal in an infinitely large space, and is used in deriving the L-G-K and K-G-T model. In this solution, the crystal shape is a paraboloid in 3D ($(r/R)^2=-2(y/R)$). In the study, the cross section of the paraboloid is plotted in Fig.\ \ref{fig:tip_morphology}a as the dashed line. The dashed line overlaps with the contour curves near $x/R=0$, demonstrating the consistency of the simulation result with Ivantsov's solution. 

The transverse sections of the simulated tip are shown in Fig.\ \ref{fig:tip_morphology}b using solid circles, with the colors indicating different sections. The interval between sections along the Z direction is 0.2 $\rm \mu m$. The distance between the first section (black solid circles) and dendrite tip is also 0.2 $\rm \mu m$. Obviously, the section near the tip is nearly circular and then gradually deviates from circular when far away from dendrite tip, indicating the shape's transition from axisymmtric to nonaxisymmtric. This phenomenon is consistent with those found in Fig.\ \ref{fig:steady_state_r}b and Fig.\ \ref{fig:tip_morphology}a. 

The solid curves in Fig.\ \ref{fig:tip_morphology}b are shapes of the Fourier decomposition with different $z^{'}$: 
\begin{equation}
    r^2(z^{'},\phi)=\sum_n A_n(z^{'})cos(4n\phi)
    \label{eq:fourier}
\end{equation}
where ($r,z^{'},\phi$) are cylindrical coordinates with the origin point being the dendrite tip, and $z^{'}$ is $z/R$, where $z$ is the distance from the section to the dendrite tip. $A_n(z^{'})$ is factor linked to the degree of surface energy anisotropy ($\varepsilon$). In this study, $A_0(z^{'})$ is $-2z^{'}$, $A_1(z^{'})$ is $0.1408|z^{'}|^{1.603}$ corresponding to $\varepsilon=0.03$ \cite{karma1998quantitative}, and $A_n(z^{'}) (n>1)$ is ignored as their contributions to the shapes of the Fourier decomposition is much smaller \cite{karma1998quantitative}. Apparently, the simulated contours match well with the shapes of Fourier decomposition, indicating the tip morphology is dominated by the fourfold symmetry mode. This result is also the same as the simulation result of the phase field model \cite{karma1998quantitative}, where the anisotropy surface energy is included in the equation of free energy, but the computational cost is usually much higher. Furthermore, this agreement also verifies the speculation of $\varepsilon$ for the current model in Section.\ \ref{S:L-G-K}.

\subsubsection{Mesh convergence tests}

\begin{figure}[htp]
    \centering
    \includegraphics[width=0.55\linewidth]{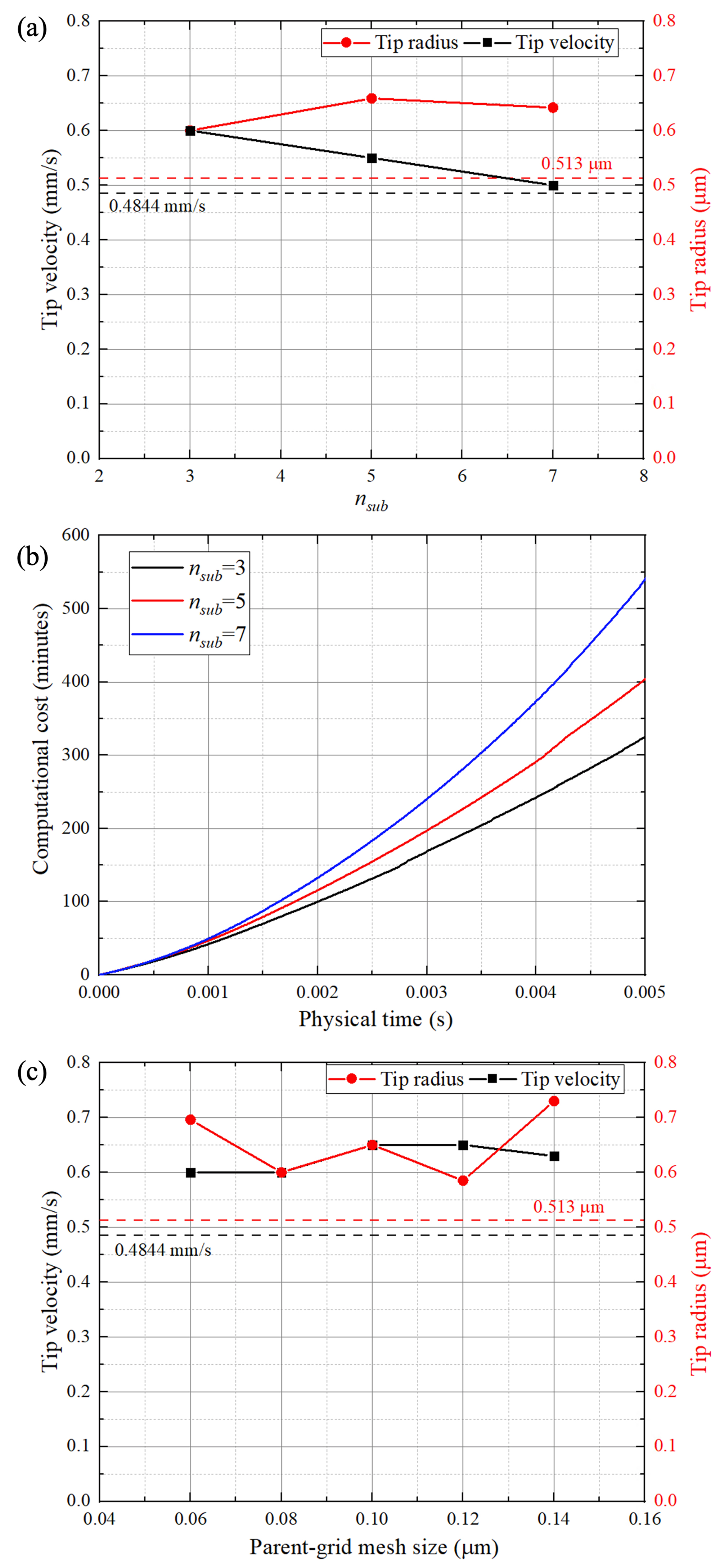}
    \caption{(a) The simulated tip velocities and radii with different $n_{sub}$, the dashed lines are the calculated tip velocity (black) and tip radius (red) using the K-G-T model with $\varepsilon$ equal to 0.03; (b) the computational cost with different $n_{sub}$; (c) The simulated tip velocities and radii with different $d_{CELL}$, the dashed lines are the calculated tip velocity (black) and tip radius (red) using the K-G-T model with $\varepsilon$ equal to 0.03.}
    \label{fig:n_sub_mesh_study}
\end{figure}

In the present model, a parent-gird cell in the interfacial region is further divided into $n_{sub}^3$ child-grid cells, and $n_{sub}$ is set to be odd number. To investigate the impact of the $n_{sub}$ on the model performance, simulations with $n_{sub}$ being 3, 5, 7 are conducted. In these simulations, $\Delta T$ is 6 K, $d_{CELL}$ is 0.08 $\rm \mu m$ and other settings are the same as those in Section.\ \ref{S:L-G-K}. As shown in Fig.\ \ref{fig:n_sub_mesh_study}a, the simulated tip velocity tends to be closer to the prediction of the K-G-T model ($\varepsilon=0.03$) when using a finer child grid, while the simulated tip radii show almost no improvement. There are two causes for these phenomena: (1) the finer child mesh can enhance the model's ability of capturing the movement of the solid-liquid interface in the parent-grid cell; (2) the curvature is only evaluated in the parent grid, which simplifies the curvature calculation but limits the improvement due to the refined child grid.

However, the computation cost would increase due to the refined grid as shown in Fig.\ \ref{fig:n_sub_mesh_study}b. As the child grid only exists in the interfacial region, the discrepancy is small when the dendrite is very small, and increases gradually as the dendrite grows up. Additionally, a sharper increase of the computation cost is obvious for simulation using a bigger $n_{sub}$. From these two figures, it is clear that a lower computational cost and acceptable simulation result can be obtained with $n_{sub}=3$. Thus, it is adopted in the current study.

To investigate the impact of the parent-grid mesh size, the simulation of dendrite growth at the melt of $\Delta T=6$ K is conducted in the same calculation domain with different parent-grid mesh sizes (0.06 $\rm \mu m$-0.14 $\rm \mu m$). The simulated tip velocities and radii are presented in Fig.\ \ref{fig:n_sub_mesh_study}c. Obviously, the simulated velocities and radii are relatively consistent using different mesh sizes, indicating that the model can obtain stable output when changing the mesh size. Nonetheless, it is important and natural to note that the accuracy would decrease with the further increase of the mesh size due to the resulted coarser solid-liquid interface. As for the computational cost, as we all know, the finer the mesh, the higher the computational cost.

\subsection{Model validation}
\begin{figure}[htp]
    \centering
    \includegraphics[width=0.6\linewidth]{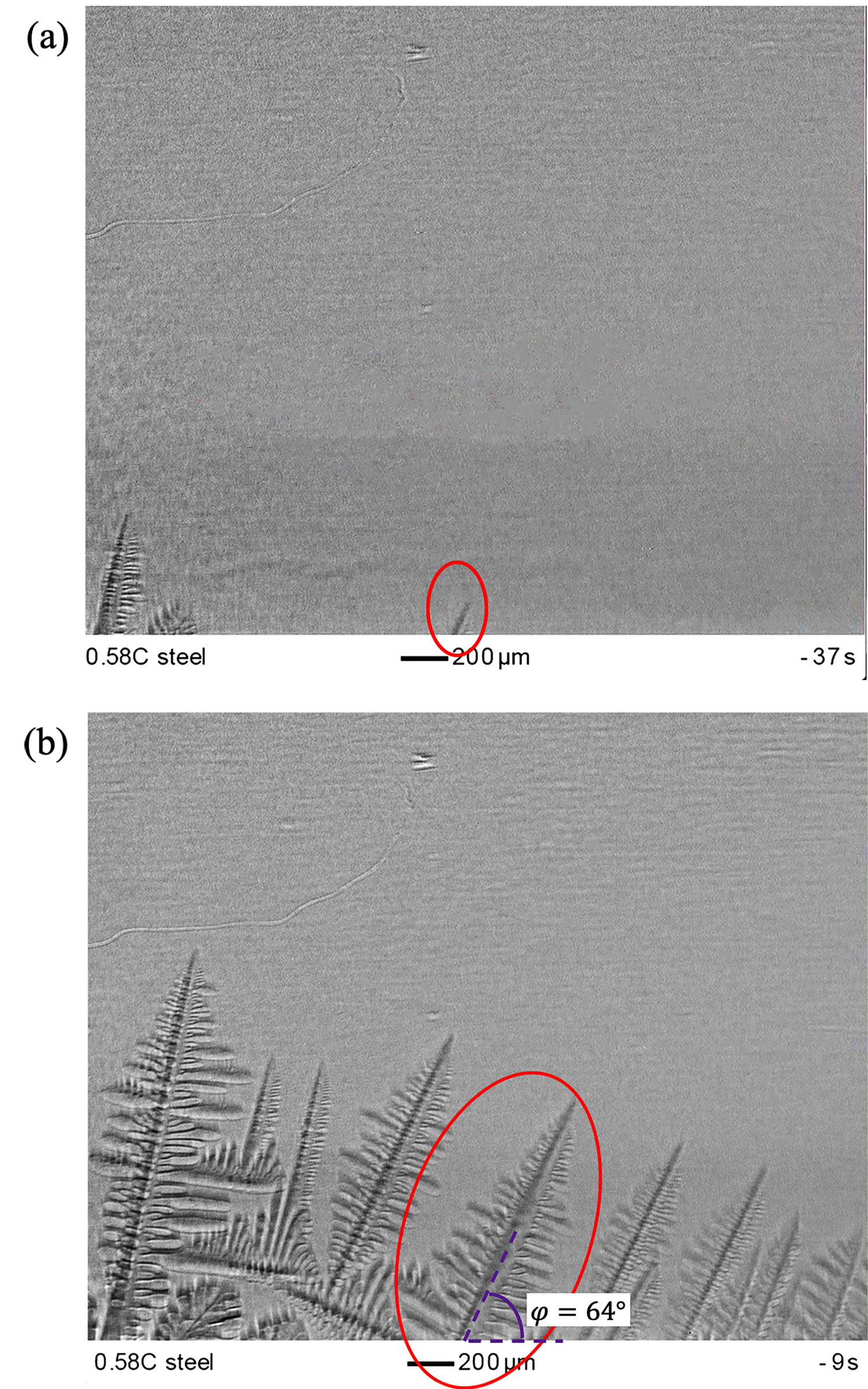}
    \caption{The experimental observation using the x-ray imaging technique at (a) t=-37 s and (b) t=-9 s (reprint from \cite{yasuda2019dendrite} under the CC BY license). The evolution of the length of the dendritic trunk highlighted by the red ellipse is extracted and compared with the simulation results.}
    \label{fig:experiement}
\end{figure}

\begin{figure}[htp]
    \centering
    \includegraphics[width=0.6\linewidth]{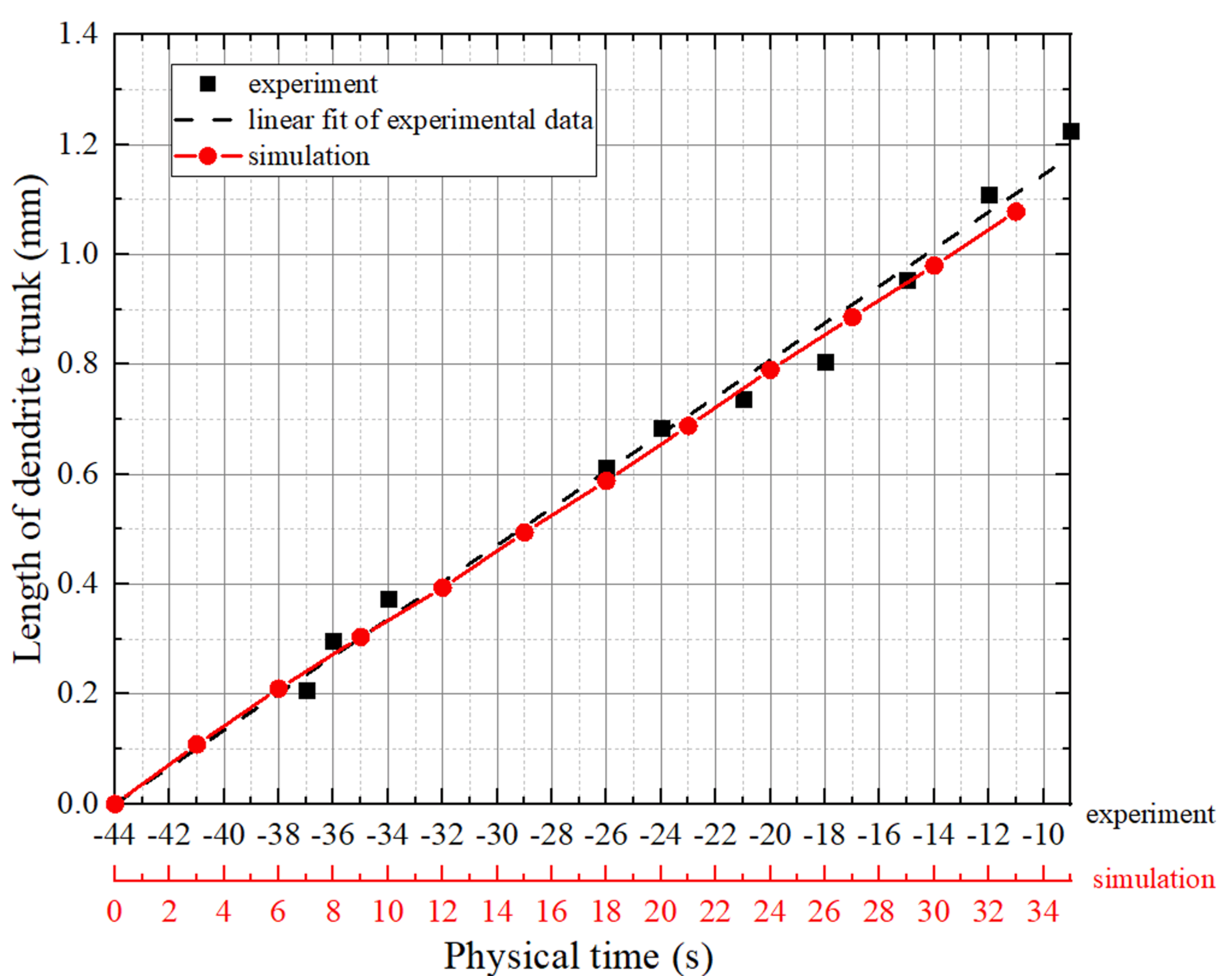}
   \caption{The evolution of the dendrite length in the experiment and simulation.}
   \label{fig:dendrite_length}
\end{figure}

\citet{yasuda2019dendrite} observed the dendrite growth process in Fe-0.58 wt.\%C alloy melt using the x-ray imaging technique and presented the process in the form of video, which is used to directly validate our model. In the video, from $t=-37$ s to $t=-9$ s and before the $\delta \rightarrow \gamma$ phase transformation occurs, the dendrite (red ellipse in Fig.\ \ref{fig:experiement}, and $\varphi =64^{\circ}$ ) grows steadily from 0.20 mm to 1.15 mm. In this study, the evolution of the dendrite trunk length is extracted and plotted in Fig.\ \ref{fig:dendrite_length}, from which we can see that growth velocity ($V_m$) is relatively constant around 0.0337 mm/s. As described in the reference \cite{yasuda2019dendrite}, during the dendrite growth process, the cooling rate ($R_c$) is 0.17 K/s and there is a temperature gradient ($G_T$) along the vertical direction. However, the exact temperature gradient is not given in their article. Nevertheless, since the growth velocity ($V_m$) is relatively constant, it is reasonable to consider the undercooling at the dendrite tip to be invariable, which then leads to the equation $ V_m \cdot sin(\varphi)\cdot G_T = R_c$. Therefore, the temperature gradient is approximated to be $5.6\times 10^3$ K/m.

The simulation is conducted on a $80\times 250\times 300$ mesh with the parent-grid cell size $d_{CELL}=4$ $\rm \mu m$ and the child-grid cell size $d_{cell}=d_{CELL}/3$. The initial temperature at the bottom of the simulation domain is 1762.75 K, i.e. the undercooling is 2 K, which is determined using the measured growth velocity ($V_m$) based on the K-G-T model. Additionally, the cooling rate and the temperature gradient are the same as those in the experiment, and the time step size is calculated to be 2 ms. At the beginning, a dendrite seed with the same crystallographic orientation as that of the dendrite highlighted in Fig.\ \ref{fig:experiement}b is placed on cell (40, 150, 0). During the simulation, the dendrite is not allowed to cross the domain boundary in this model. This simulation is carried out on a desktop equipped with Intel core i7-9700 and costs about 24 hours. 

\begin{figure}[htp]
    \centering
    \includegraphics[width=\linewidth]{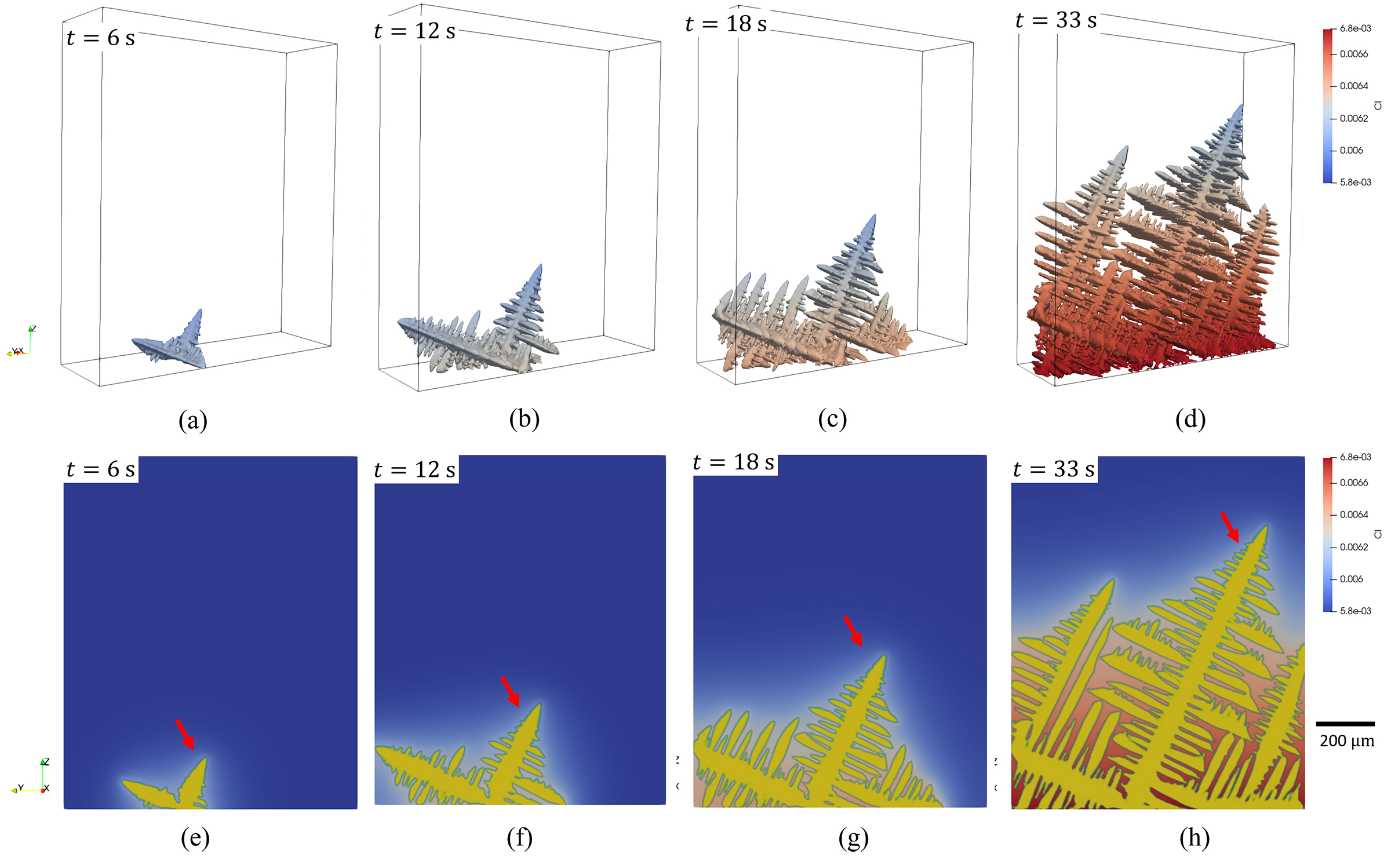}
    \caption{The simulated microstructure at (a) $t=6$ s, (b) $t=12$ s, (c) $t=18$ s and (d) $t=33$ s colored by the liquid solute concentration in the interfacial cells. The cross section of the simulation domain at (e) $t=6$ s, (f) $t=12$ s, (g) $t=18$ s and (h) $t=33$ s, in which the carbon concentration field in the remaining liquid is shown accompanied by the dendrite structure (in yellow).}
    \label{fig:simulation_validation}
\end{figure}

The evolution process of the dendrite is presented in Fig.\ \ref{fig:simulation_validation}. The dendrite exhibits a branchless shape at the very beginning and then develops its secondary arms (Fig.\ \ref{fig:simulation_validation}a, b, e and f). Afterwards, the growth of the dendrite trunk at the left side is blocked by the domain boundary. The dendrite trunk stop growing longer. Some secondary arms on it that are parallel to the dendritic trunk pointed by the red arrow develop the tertiary arms (see Fig.\ \ref{fig:simulation_validation}c,d, g and h), resulting in the shape of a new dendrite trunk. In Yasuda's video \cite{yasuda2019dendrite}, although there is no boundary restricting the dendrites' growth, the left branch of the dendrite (highlighted by red circle) is blocked by the neighboring dendrite. In this branch, one of the secondary dendrite arms develop the tertiary arms, and finally become a new dendrite trunk (see Fig.\ \ref{fig:experiement}b), which is similar to the phenomenon in our simulation. Besides, as the cross-section views in Fig.\ \ref{fig:simulation_validation}e, f, g and h show, when the dendritic trunk grows towards the top of the domain, dendritic arms at the bottom coarsen continuously. These phenomena are also observed in Yasuda's video \cite{yasuda2019dendrite}. Furthermore, the length of the dendritic trunk (pointed by a red arrow in Fig.\ \ref{fig:simulation_validation}) at different times are also extracted and plotted in Fig.\ \ref{fig:dendrite_length}.  Obviously, the simulation result is in good agreement with the experimental data. According to the growth kinetics represented by Eqs.\ \ref{eq:dFs}, \ref{eq:cleq} and \ref{eq:kappa}, the accurate evaluation of the growth depends on the evaluation of the interface curvature, which needs the accurate capture of the solid-liquid interface. Therefore, the good agreement in the trunk growth curve also indicates the good capture of the solid-liquid interface in the present model.

\subsection{Dendrite growth in additive manufacturing}
In additive manufacturing, especially the Electron Beam Selective Melting (EBSM) of Ni-based superalloy, the alloy solidifies with a significant segregation in the inter-dendritic regions despite of the extremely high temperature gradient ($\approx 10^5-10^7$ K/m) and cooling rate ($\approx 10^3-10^5$ K/s). This process usually comes with the formation of precipitates and the incubation and propagation of hot cracks, which would affect the mechanical performance of the as-built products. Simulating the dendrite growth process during the solidification process would help us understand the physical mechanism, and then provide guidance on how to tailor the microstructure evolution during the AM process.

\begin{table}[htp]
    \centering
    \caption{The material properties of Inconel 718 used in the thermal-fluid model}
 
    \begin{tabular}{c c}
    \hline
        Material properties & Value \\
        \hline
        Density & 8190 $\rm kg/m^3$\\
        Liqudus temperature & 1609 K\\
        Solidus temperature & 1523 K\\
        Latent heat of fusion & $2.1\times 10^5$ J/kg\\
        Thermal conductivity & 29 $\rm W/(m\cdot K)$ \\
        Specific heat & 720 $\rm J/(kg\cdot K)$\\
        Surface tension coefficient & 1.882 N/m\\
        Temperature sensitivity of surface tension coefficient & 0.0001 \\
        Boiling temperature & 3005 K\\
        Latent heat of evaporation & $6.31\times 10^6$ J/kg \\
        Viscosity & 0.0072 $\rm Pa\cdot s$\\
        \hline
    \end{tabular}
    
       \label{tab:718_tf}
\end{table}

\begin{figure}[htp]
    \centering
    \includegraphics[width=\linewidth]{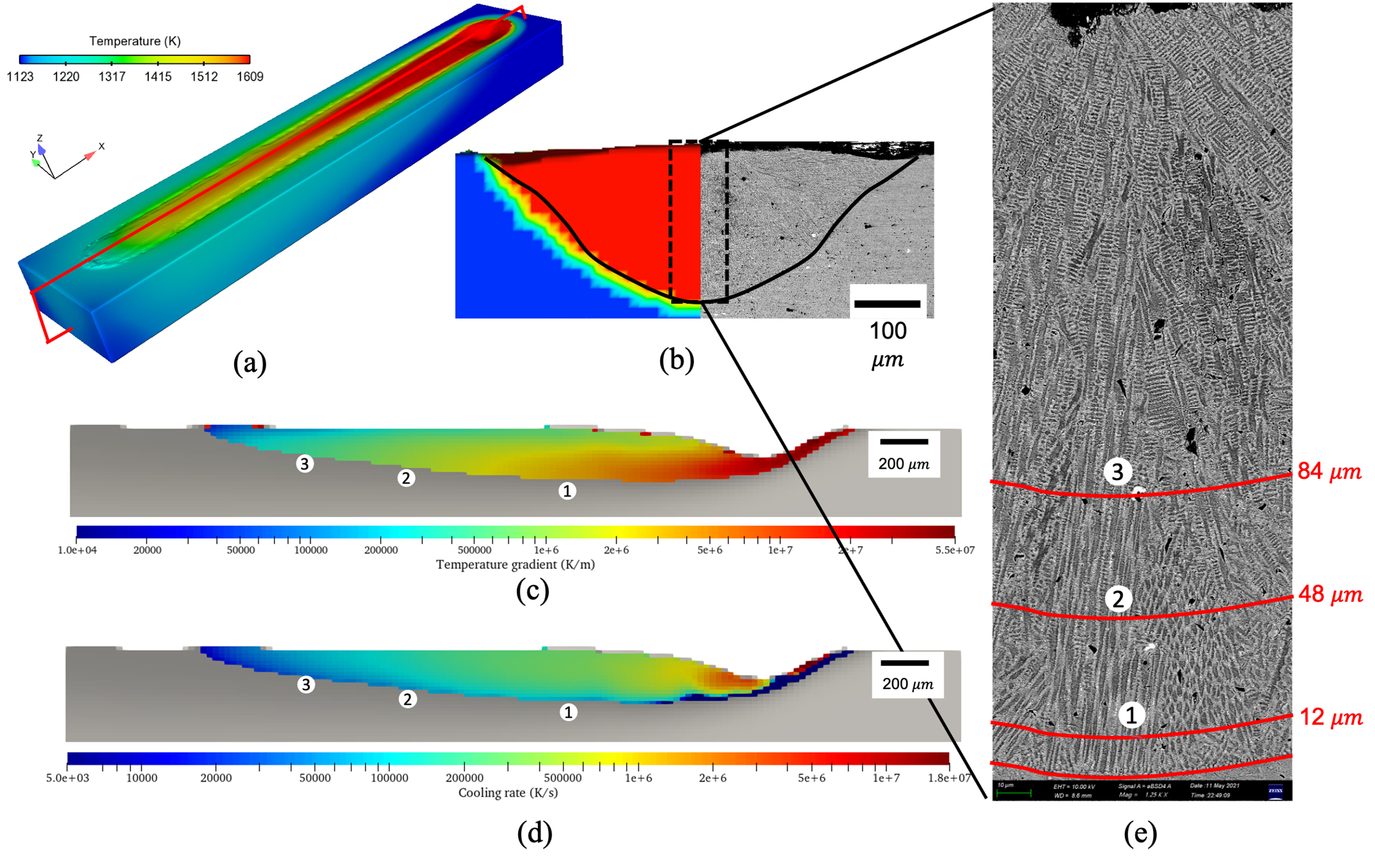}
    \caption{(a) Simulated track using the thermal-fluid model. The red rectangle indicates the center plane. (b) The comparison of the cross sections obtained by the thermal-fluid simulation (left) and the experiment (right), the region in red indicates the melted region. (c) The temperature gradient field at the center plane. (d) The cooling rate field at the center plane. In Fig.(c) and (d), only the data at the melting pool region ($T \geq 1609$ K) is shown, while the other regions are in gray. (e) The magnification of the microstructure in the solidified melting-pool region.}
    \label{fig:AM_EX_single_track}
\end{figure}

\begin{figure}[htp]
    \centering
    \includegraphics[width=0.6\linewidth]{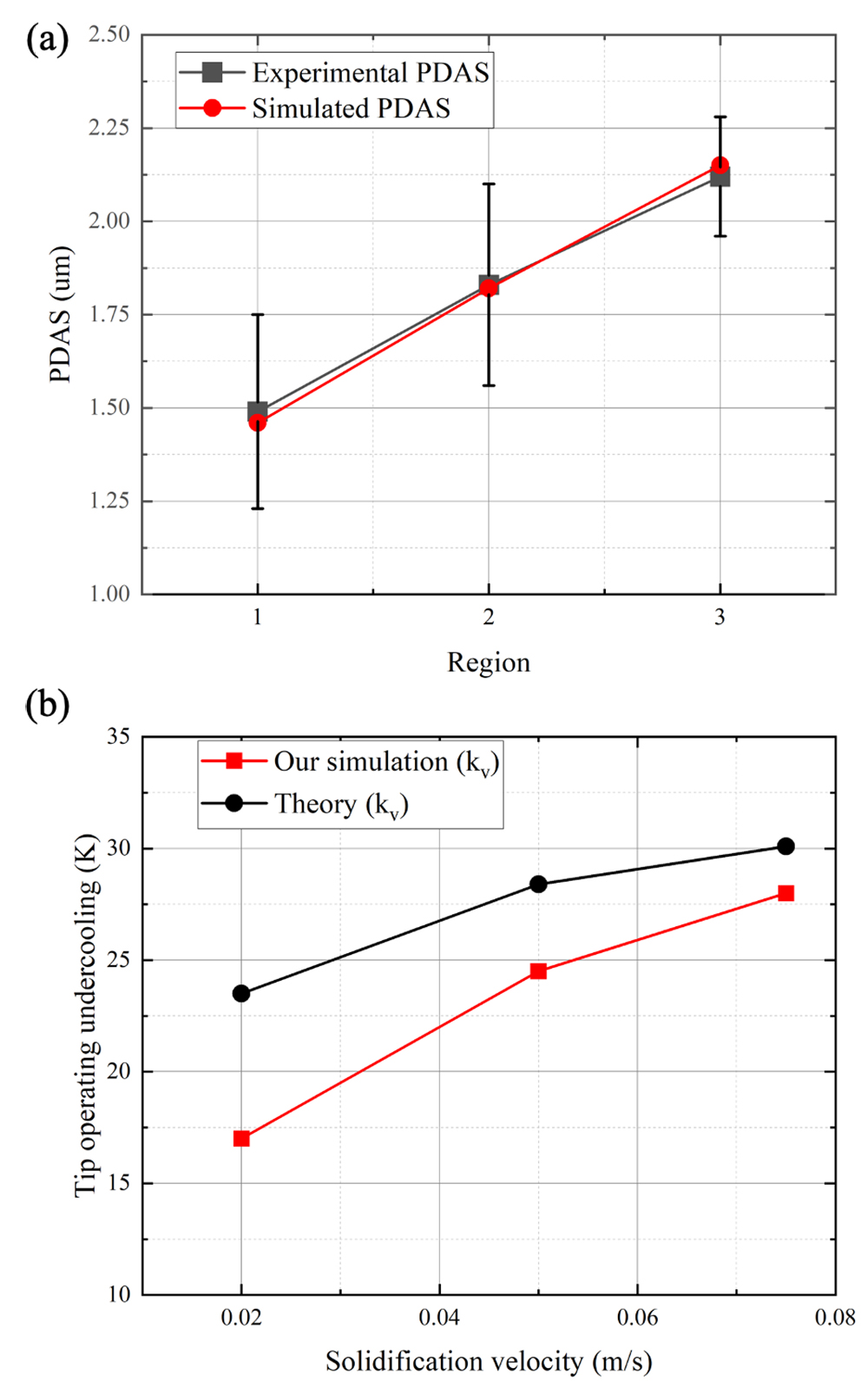}
    \caption{(a) The comparison of simulated and measured PDAS; (b) The comparison of the simulated tip undercooling and the theoretical value (with $V$-dependent $k_v$) extracted from \cite{ghosh2018simulation}}
    \label{fig:PDAS_value}
\end{figure}

Single track is the fundamental unit of additive manufacturing. In this study, to demonstrate the model's validity in AM condition, a single-track scanning experiment on an Inconel 718 substrate is carried out in a self-developed EBSM machine. The substrate was preheated at the beginning with the defocused electron beam. The scanning with focused electron beam started when the temperature of the substrate reached 1123 K. The beam power is 400 W, and the scanning speed is 0.5 m/s. The sample was cut along the plane perpendicular to the scanning direction in order to show the cross section of the track. The cross section was polished and then etched for microstructure observation in a scanning electron microscope (SEM). Meanwhile, energy dispersive X-ray spectroscopy (EDS) analysis was conducted to characterize the chemical concentration field. The single track has a width of 702 $\rm \mu m$ and a depth of 219 $\rm \mu m$. In Fig.\ \ref{fig:AM_EX_single_track}e, three regions that are located near the center plane are chosen for the comparison with the simulation results using the presented dendrite growth model. The distance between the regions and the bottom of the melting pool are 12 $\rm \mu m$, 48 $\rm \mu m$ and 84 $\rm \mu m$, respectively. In the melting pool, the dendrites prefer to grow towards the center of the melting pool, making the dendrite structure at the upper part of the melting pool extremely complex (see Fig.\ \ref{fig:AM_EX_single_track}e). Therefore, the selected regions are all at the lower part of the melting pool, where the critical indicator, i.e., the primary dendrite arm spacing (PDAS), is measurable. The PDASs in these regions are measured and plotted in Fig.\ \ref{fig:PDAS_value}a. 

A thermal-fluid model developed by \citet{yan2017multi} is used to reproduce the melting and solidification process. In this model, the major factors including the fluid flow, Marangoni effect, metal evaporation, and recoil pressure are incorporated, so that a reliable thermal field can be obtained, which has been validated in our previous work \cite{2019Pore,wang2020evaporation}. In this simulation, the initial temperature of the substrate was set as 1123 K instead of modelling the preheating process. The other scanning parameters are exactly the same as those 
used in the experiment. The beam diameter was set as 700 $\rm \mu m$ by calibrating against the experimental result. The material properties are listed in Table\ \ref{tab:718_tf}. The simulation result is shown in Fig.\ \ref{fig:AM_EX_single_track}. Obviously, the comparison of the cross sections perpendicular to the scanning direction in Fig.\ref{fig:AM_EX_single_track}b indicates that the scanning process is well reproduced by the thermal-fluid model. According to the distances between the selected regions and the bottom of the melting pool (Fig.\ \ref{fig:AM_EX_single_track}e), the corresponding solidification conditions (temperature gradient and cooling rate) are extracted from the thermal-fluid simulation (Fig.\ \ref{fig:AM_EX_single_track}c and d). The detailed data are shown in Table\ \ref{tab:AM_parameter}. These parameters are then used in simulating the dendrite growth in the three regions, respectively.

\begin{table}[htp]
    \centering
    \caption{The simulated solidification condition at the three regions}
    \begin{tabular}{c c c c}
    \hline
        Region &  1 & 2 & 3\\
        \hline
         Temperature gradient (K/m)& $5\times 10^6$ & $1\times 10^6$ & $4\times 10^5$\\
         Cooling rate (K/s) & $1\times 10^5$ & $5\times 10^4$ & $3\times 10^4$\\
         Solidification velocity (m/s)& 0.02 & 0.05 & 0.075\\
         Partition coefficient & 0.496 & 0.519 & 0.536\\
         Liqudus slope ($\rm K/(wt. \%$)) & -10.505 &-10.531 &-10.564\\
         \hline
    \end{tabular}
    \label{tab:AM_parameter}
\end{table}

In this study, the dendrite growths in the three regions are carried out using the temperature gradients and cooling rates listed in Table\ \ref{tab:AM_parameter}. Note that the temperature gradient and cooling rate are constant in the dendrite growth model. Since Nb is one of the most important segregation elements in Inconel 718, and the Nb-rich precipitates in the inter-dendritic regions significantly affect the mechanical performance of the as-built products, the material is treated as Ni-Nb binary alloy following \cite{nie2014numerical,ghosh2017primary,xiao2019multi}. The concentration of Nb is 5 wt.\%. The detailed physical properties are listed in Table.\ \ref{tab:NiNb_alloy}. The simulation domain is a $60\times 60\times 100$ mesh with $d_{CELL}=0.2$ $\rm \mu m$ and $n_{sub}=3$. Before the simulations start, a number of dendrite seeds (about 420) with the same crystalline orientation are randomly placed on the bottom of the simulation domain. The distance between these dendrite seeds is restricted below 0.6 $\rm \mu m$. The initial temperature at the bottom of the simulation domain is set as the liquidus temperature of Inconel 718. 

 \begin{table}[htp]
    \centering
        \caption{Critical physical properties of Ni-Nb alloy}
    \begin{tabular}{c c}
    \hline
    Properties & Values\\
    \hline
       Solute diffusion coefficient in the liquid ($D_l$) &  $3\times 10^{-9}$  $\rm m^2/s$\\
        Partition coefficient at equilibrium state ($k$) & 0.48\\
        Liquidus slope at equilibrium state ($m_l$) & -10.5 $\rm K/(wt. \%$)\\
        Gibbs-Thomson coefficient ($\Gamma$) & $3.65\times 10^{-7}$ $\rm m\cdot K$ \\
        \hline
    \end{tabular}
    \label{tab:NiNb_alloy}
\end{table}

\begin{figure}[htp]
    \centering
    \includegraphics[width=\linewidth]{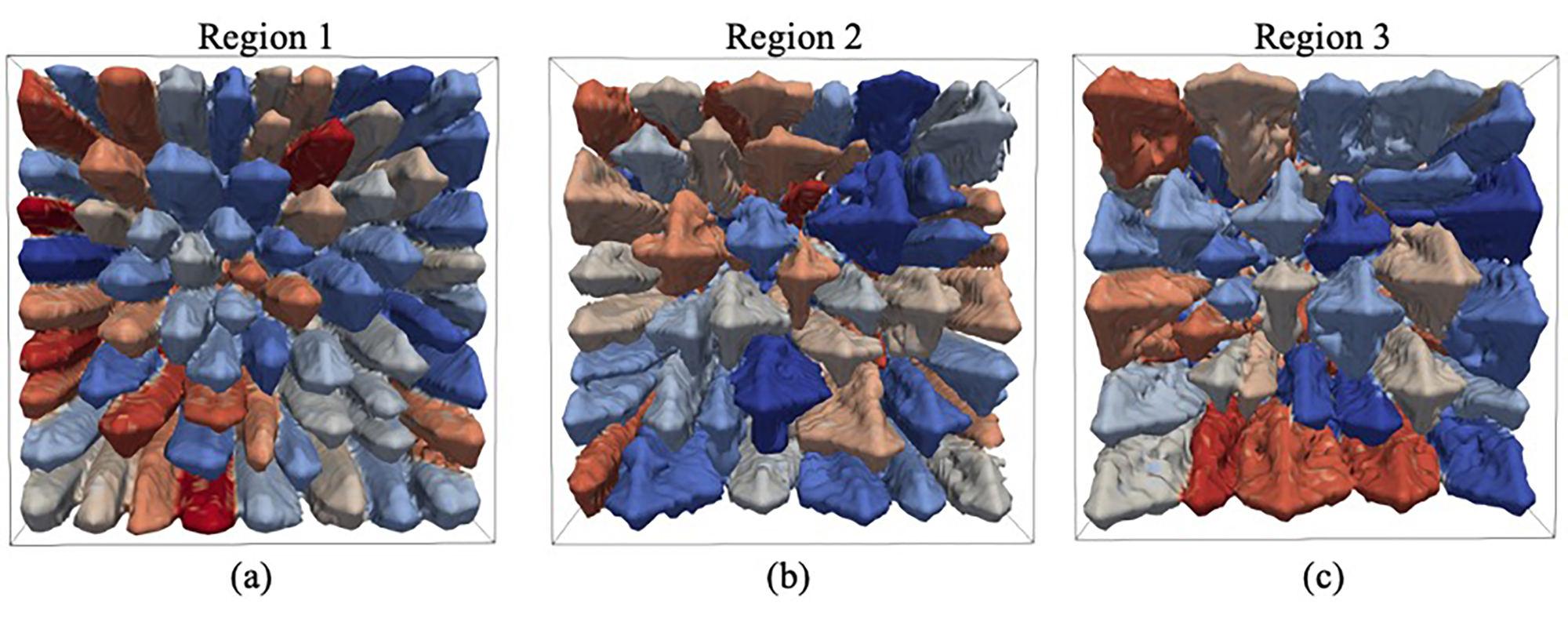}
    \caption{The simulated dendrite structure observed from the top: (a) region 1; (b) region 2; (c) region 3. The color used in these figures are for distinguishing dendrites.}
    \label{fig:AM_dendrite}
\end{figure}

As shown in Table\ \ref{tab:AM_parameter}, the solidification velocities are high so that the solute trapping cannot be ignored. In this study, the solidification velocity ($V$)-dependent solidification parameters (partition coefficient $k_v$ and the liquidus slope $m_v$) are calculated according to Eq.\ \ref{eq:aziz} (Aziz's model \cite{aziz1988continuous}) and Eq.\ \ref{eq:mv} \cite{kurz1989fundamental}, and adopted in the simulations. In Eq.\ \ref{eq:aziz}, $V_{D}$ is the interface diffusion velocity, which is material-dependent. Up to date, there is no precise measurement of $V_D$ for Inconel 718. \citet{ghosh2018simulation} fitted the simulated solidification velocity-partition coefficient relation with Aziz's model, thereby determining $V_D=0.62$ m/s. This result is consistent with the deduction ($V_D<1$ m/s) made by \citet{tao2019crystal} based on their experimental results. Therefore, in the study, $V_D=0.62$ m/s, and then the corresponding $k_v$ and $m_v$ for the three regions are calculated (see Table\ \ref{tab:AM_parameter}) and adopted in the dendrite growth simulations.

\begin{equation}
    k_v=\frac{k_e+V/V_{D}}{1+V/V_{D}}
    \label{eq:aziz}
\end{equation}

\begin{equation}
    m_v=m_l(1+\frac{k_e-k_v[1-ln(k_v/k_e)]}{1-k_e})
    \label{eq:mv}
\end{equation}

 Fig.\ \ref{fig:AM_dendrite} shows the simulated dendrite structures when the tip undercooling is under the steady state. Obviously, the number of dendrites decreases from region 1 to region 3. It is critical to note that the number of dendrites is much smaller than that of the dendrite seeds placed on the domain bottom, which is due to the growth competition among dendrites. In this study, the simulated PDAS are calculated by $\lambda_1=\sqrt{A/N_d}$ ($A$ is the area of the XY cross section, and $N_d$ is the number of remaining dendrites observed from the domain top). As shown in Fig.\ \ref{fig:PDAS_value}a, the simulated PDASs are quite close to the measured results. The tip undercoolings obtained when the dendrite growth reaches the steady state are plotted in Fig.\ \ref{fig:PDAS_value}b. The theoretical values (considering the $V$-dependent $k_v$) calculated by \cite{ghosh2018simulation} are also plotted for comparison. Both the theoretical values and the simulated values indicate that the higher solidification velocity comes with a higher tip undercooling. Although the simulated results are all lower than the theoretical predictions, the differences between the simulation results and the theoretical predictions are not that big ($V=0.02$ m/s: 6.5 K or 27.7\%; $V=0.05$ m/s: 3.9 K or 13.7\% and $V=0.075$ m/s: 2.1 K or 7\%). That is, the simulation results are fairly close to the theoretical predictions. The differences are mainly the impact of the discretization in the model. Especially, since $d_{CELL}=0.2$ $\rm \mu m$, the temperature change $\delta T$ between neighboring cells along the $Z$ direction are 1 K ($V=0.02$ m/s), 0.2 K ($V=0.05$ m/s) and 0.08 K ($V=0.075$ m/s), respectively. Obviously, the smallest $\delta T$ in region 3 results in the smallest difference as shown in Fig.\ \ref{fig:PDAS_value}b. Additionally, in comparison with the phase field model in \cite{ghosh2018simulation}, the present model shows a much better consistency with the theoretical prediction when the solidification velocity is higher than 0.03 m/s. 

\begin{figure}[htp]
    \centering
    \includegraphics[width=\linewidth]{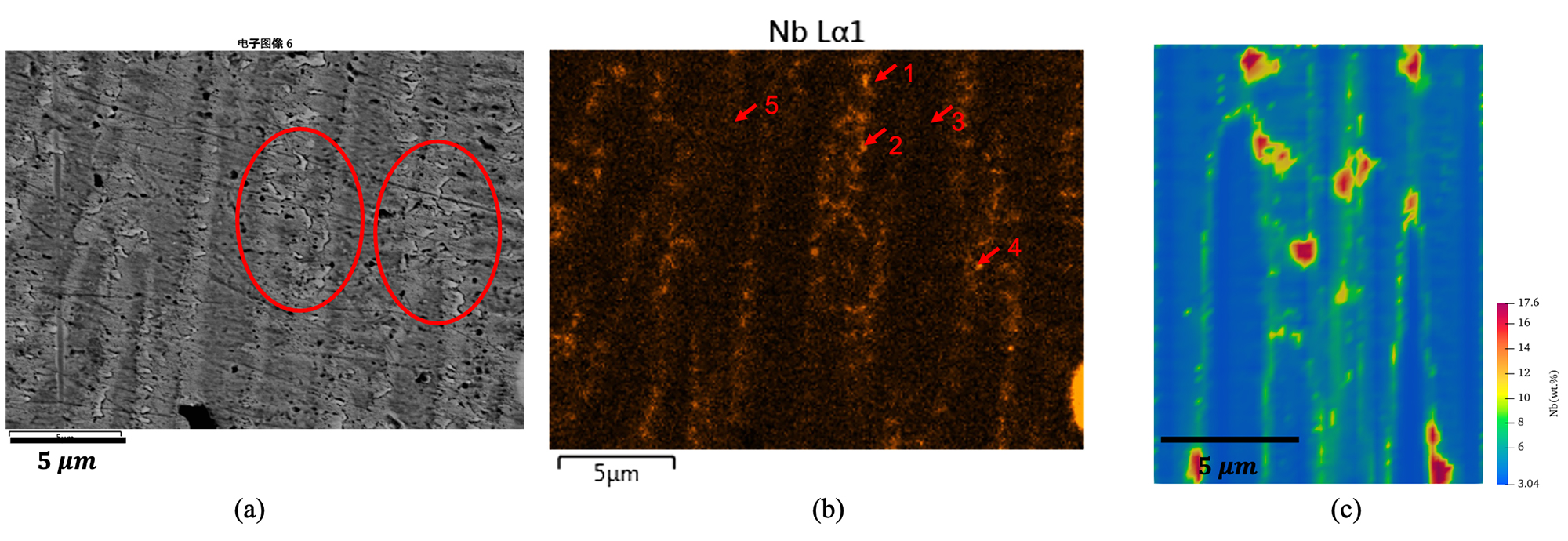}
    \caption{(a) The microstructure at the region 3. (b) The Nb concentration field of the region shown in (a) (measured by EDS). (c) The simulated Nb concentration field.}
    \label{fig:Nb_concentration}
\end{figure}

\begin{table}[htp]
    \centering
        \caption{The measured Nb concentration at the points shown in Fig.\ \ref{fig:Nb_concentration}b}
    \begin{tabular}{c c c c c c c}
    \hline
       point  & 1 & 2& 3& 4&5 \\
       \hline
        Nb (wt.\%) & 27.1&21.91&2.57&26.56&3.73\\
        \hline
    \end{tabular}
    \label{tab:Nbpt}
\end{table}

As Nb-rich precipitates significantly affect the mechanical performance of the as-built parts, the simulated Nb distribution is compared with the measured result. Firstly, both the simulation result and experimental observation show that Nb prefers to concentrate at places where the dendrites compete for growth space fiercely. For example, the regions highlighted by red circles in Fig.\ \ref{fig:Nb_concentration}a, and the middle region in Fig.\ \ref{fig:Nb_concentration}c. At these regions, the Nb concentrations are significantly higher than those at ordinary inter-dendritic regions. Secondly, the Nb concentrations at representative points (see Fig.\ \ref{fig:Nb_concentration}b) are listed in Table\ \ref{tab:Nbpt}. Point 3 and 5 are located at the dendrite trunks, the Nb concentrations are similar to the simulation results at the dendrite trunks. Thirdly, in the simulation result and the experimental result, there are some small regions with very high Nb concentration, such point 1, 2 and 4 in Fig.\ \ref{fig:Nb_concentration}b, and red regions in Fig.\ \ref{fig:Nb_concentration}c. These regions are all located at the fierce-competition places. Obviously, the measured Nb concentration in point 1, 2 and 4 are all higher than the maximum value in the simulation result, which is due to the further element concentration caused by the formation of Nb-rich precipitates (see the corresponding points in Fig.\ \ref{fig:Nb_concentration}a). Detailed investigation into this issue is not in the scope of the present work.

\section{Conclusion}
\label{S:conclusion}

In the present work, a multi-grid Cellular Automaton model is developed for simulating the dendrite growth. In this model, the calculation domain is firstly discretized into cubic cells (parent grid), and then cells at the solid-liquid interfacial regions are further divided into smaller cubic cells (child grid), where the decentered octahedron growth algorithm is applied. The model calculates the interface movement using the solute equilibrium method on the parent grid. The movement is then transferred to the child grid to drive the growth and capture of envelopes, thereby updating the solid fraction in the child grid and the parent grid. 

The model's ability of simulating the growth of dendrites with arbitrary crystalline orientations is firstly verified with a multi-dendrite growth simulation. Single-dendrite-growth simulations of Fe-0.6 wt.\%C with different undercoolings are conducted. The simulated tip velocities and radii agree well with the predictions of the K-G-T model(a simplified L-G-K model). Meanwhile, the degree of surface energy anisotropy that is implicitly incorporated in the current model is deduced and approximately equals to 0.03. Additionally, the steady state tip morphology is analyzed in detail, the shape of the tip is demonstrated to agree well with the Ivantsov solution, being nonaxisymmetric and deviating from the paraboloid in the fourfold symmetry mode. Afterwards, the model is validated against a published x-ray imaging observation. Additionally, the model shows comparable accuracy with the CA model in the first category, and is the first 3D CA-based dendrite growth model, which belongs to the second category and uses decentered octahedron growth algorithm, to be comprehensively and quantitatively verified accurate.

At last, the model's validity in additive manufacturing is validated against a single-track experiment. The single-track scanning process is firstly simulated using a thermal-fluid model. The extracted solidification conditions at different depths are then used for modeling the dendrite growth processes. The simulated PDASs and Nb concentration show fairly good consistency with the experimental results. The tip undercoolings are compared with the theoretical values calculated by \cite{ghosh2018simulation}, and show a reasonable agreement. Therefore, this model would be a promising method for investigating critical physical mechanisms including the nucleation, the formation of shrinkage porosity, the incubation and propagation of hot cracking and the formation of second phase, which are related to the dendrite growth process.

\section*{Acknowledgement}
Yefeng Yu and Prof. Feng Lin thank the financial support of the National Key R\&D Program of China (2017YFB1103303). Yefeng Yu and Dr. Wentao Yan acknowledge the support of Singapore Ministry of Education Academic Research Fund Tier 1. 





\bibliographystyle{elsarticle-num-names}
\bibliography{sample.bib}







\end{document}